\documentclass[twocolumn]{IEEEtran}
\usepackage{multicol,lipsum}
\usepackage{authblk}
\usepackage{silence}
\usepackage{placeins}
\usepackage{epsfig}
\usepackage[font=scriptsize]{caption}

\usepackage{amsmath, amsthm, amssymb}
\usepackage[table,xcdraw]{xcolor}
\usepackage{array}
\usepackage{cite}
\usepackage{amssymb}
\usepackage{color}
\usepackage{amsmath}
\usepackage{multirow}
\usepackage{dirtytalk}
\usepackage{soul}
\usepackage{hhline}
\usepackage{colortbl}
\usepackage{algorithm} 
\usepackage{algpseudocode} 
\usepackage{enumerate}
\usepackage{colortbl}
\usepackage{siunitx}
\usepackage{booktabs}
\usepackage{graphicx}
\begin{document}

\title{Dual-Stage Deeply Supervised Attention-based Convolutional Neural Networks for Mandibular Canal Segmentation in CBCT Scans}

\author[1,2]{Muhammad Usman}
\author[1]{Azka Rehman}
\author[1]{Amal Saleem}
\author[3]{Rabeea Jawaid}
\author[1]{Shi‑Sub Byon}
\author[1]{Sung‑Hyun Kim}
\author[3]{Byoung Dai Lee}
\author[1]{Byung‑il Lee}
\author[2]{Yeong‑Gil Shin}
\affil[1]{Center for Artificial Intelligence in Medicine and Imaging, HealthHub Co. Ltd., Seoul, 06524, South Korea}
\affil[2]{Seoul National University, Seoul, Republic of Korea}
\affil[3]{Division of AI and Computer Engineering, Kyonggi University, Suwon, Republic of South Korea}
\setcounter{Maxaffil}{0}

\maketitle
\begin{abstract}
Accurate segmentation of mandibular canals in lower jaws is important in dental implantology. Medical experts determine the implant position and dimensions manually from 3D CT images to avoid damaging the mandibular nerve inside the canal. In this paper, we propose a novel dual-stage deep learning-based scheme for the automatic segmentation of the mandibular canal. Particularly, we first enhance the CBCT scans by employing the novel histogram-based dynamic windowing scheme, which improves the visibility of mandibular canals. After enhancement, we design 3D deeply supervised attention U-Net architecture for localizing the volumes of interest (VOIs), which contain the mandibular canals (i.e., left and right canals). Finally, we employed the multi-scale input residual U-Net architecture (MS-R-UNet) to segment the mandibular canals using VOIs accurately. The proposed method has been rigorously evaluated on 500 scans. The results demonstrate that our technique outperforms the current state-of-the-art segmentation performance and robustness methods.
\end{abstract}

\section{Introduction}\label{sec:introduction}

Inferior alveolar nerve (IAN), also known as mandibular canal, is the most critical structures in the mandible region that supplies sensation to the lower teeth. The sensation, provided to lips and chin is via the mental nerve which passing through the mental foramen \cite{phillips2002facial}. One of a very critical steps in implant placement, third molar extraction, and various other craniofacial procedures including orthognathic surgery, is determining the position of the mandibular canal. Patients may experience aches and pain and temporary paralysis if the mandibular canal get injured \cite{lee2011impact} \cite{juodzbalys2011injury} during any of these process. Localization of the mandibular canal is important not only for diagnosis of vascular and neurogenic diseases associated with the nerve, but also for diagnosis of lesions near the mandibular canal, and planning of oral and maxillofacial procedures. Therefore, preoperative treatment planning and simulation are necessary to avoid nerve injury. The identification of exact location of can assist in achieving the planning strategy required for the task at hand \cite{gaggl2001navigational}.

One of the most frequently used three-dimensional (3D) imaging modalities for preoperative treatment planning and postoperative assessment in dentistry is Cone Beam Computed Tomography which is also known as CBCT \cite{dj7020052}. The CBCT volume is reconstructed using projection images realized from different angles with a cone-shaped beam and stored as a sequence of axial images \cite{hatcher2010operational}. A clinical replacement is multi-detector computed tomography (MDCT), but its application is limited by high radiation dose and insufficient spatial resolution. In contrast, the CBCT allows more precise imaging of hard tissues in the dentomaxillofacial area and its effective radiation dosage is lower than that of the MDCT1. In addition, CBCT is inexpensive and readily available. However, in practice, there are certain challenges associated with mandibular canal segmentation from CBCT images, such as inaccurate density and large amount of noise  \cite{angelopoulos2012comparison}.
\begin{figure*}[!h]
\centering
\centerline{\includegraphics[width=.8\textwidth]{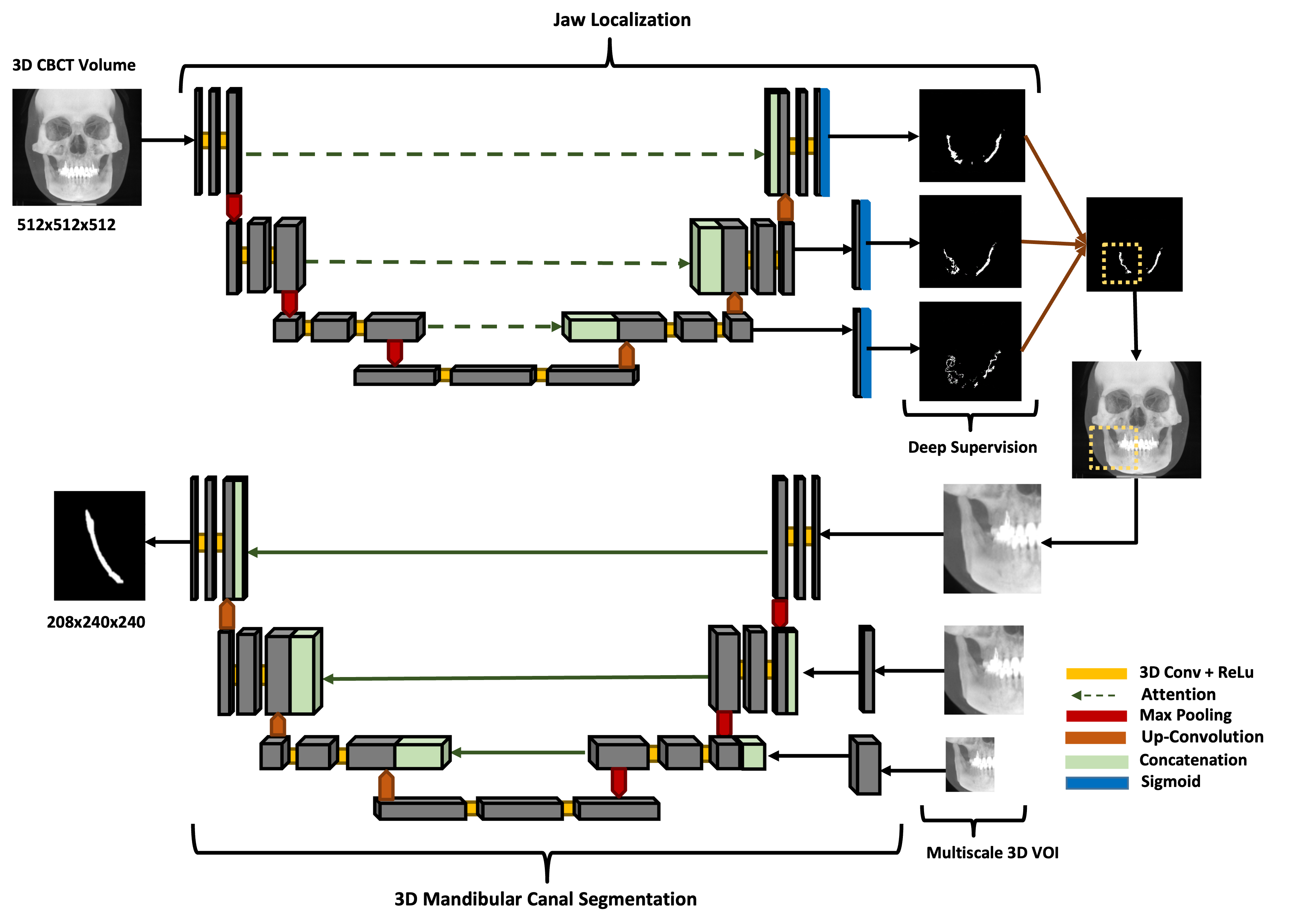}}
\caption{Proposed dual-stage scheme for mandibular canal segmentation, describing the models architectures utilized at each stage. Firstly, deeply supervised attention UNet model is used for jaw localization which coarsely segments the canal. This coarsely segmented canal is further utilized to extract VOIs which is used to produce the fine segmentation of mandibular canals (i.e., left and right canals) by employing Residual UNet with multi-scale inputs. }
\label{fig:mod}

\end{figure*}

Surgical planning and pre-surgical examination are crucial in dental clinics. One of the standard imaging tools used for such assessments and planning is a panoramic radiograph, constructed from a dental arch to provide all the relevant information in a single view. These radiographs bear disadvantages such as difficulties in determining the 3D rendering of an entire canal and connected nerves \cite{ghaeminia2011use}. One of the most common approaches for preoperative assessment is to annotate the canal in 3D images to produce the segmentation of the canal. This kind of manual annotation is a very knowledge-intensive, time-consuming, and tedious task. Thus, there is a need for a tool to assist the radiologist and reduce the burden by using automatic or semi-automatic segmentation of the canal. 


Kwak et. al. \cite{kwak2020automatic} studied different models based on 2D and 3D techniques such as on 2D SegNet, and 2D and 3D U-Nets. Their study also involved detailed pre and post-processing steps including thresholding of teeth as well as bones. Jaskari et. al.\cite{jaskari2020deep} presented an FCNN-based model to extract IAN. Dhar et. al. \cite{dhar2021automatic} used a model based on 3D UNet to segment the canal. They used pre-processing techniques to generate the center lines of the mandibular canals and used them as ground truths in the training process. Verhelst et. al. \cite{verhelst2021layered} used a patch-based technique to localize the jaw and then used the 3d UNet model to segment the canal in that ROI. Lahoud et. al. \cite{lahoud2022development} first coarsely segmented the canal, and performed fine segmentation of the canal on patches that are extracted based on the coarse segmentation. The network utilized 3D UNet with skip connections. In all the above studies, the models utilized were 3D UNet. Verhelst et al. \cite{verhelst2021layered} and Lahoud et al. \cite{lahoud2022development} utilized the localization technique and then segmented the canal. However, they used patches for fine segmentation i.e. canal is divided into multiple patches before segmentation. This reduces the visibility of the model resulting in high chances of error. Other methodologies do not take into account the varying size of the canal while training the model.
Previously work has been done in deep learning 
In this study, we propose a cascade technique to segment the mandibular canal in 3D CBCT scans. We first localize the jaw using a naively segmented canal in the form of volume of interest (VOIs). After that, we divide the canal into left and right parts and trained a multi-scale input Residual UNet model for segmenting the canal. The purpose of using multi-scale input patches is to take into account the varying size of canals.

The rest of the paper is organized as follows; In Section \ref{sec:introduction}, we present background and related work. In Section \ref{pro}, the detail on each step of our proposed method as well as the materials is described. In Section \ref{res}, we present the obtained results and comparison of our study. Finally, we analyze and discuss our work in Section \ref{disc} and conclude in Section \ref{co}.

\section{Materials and Methods}
\label{pro}
\subsection{Study Design}
The objective of this study is to design a deep-learning approach for automatic mandibular canal segmentation. The study design consists of pre-processing, model training, and post-processing each discussed in detail in the following sections. The detailed network design is discussed in section \ref{arch} Network Architecture. The network was validated on 500 scans.

\begin{figure*}[h]
\centering
\centerline{\includegraphics[width=.9\textwidth]{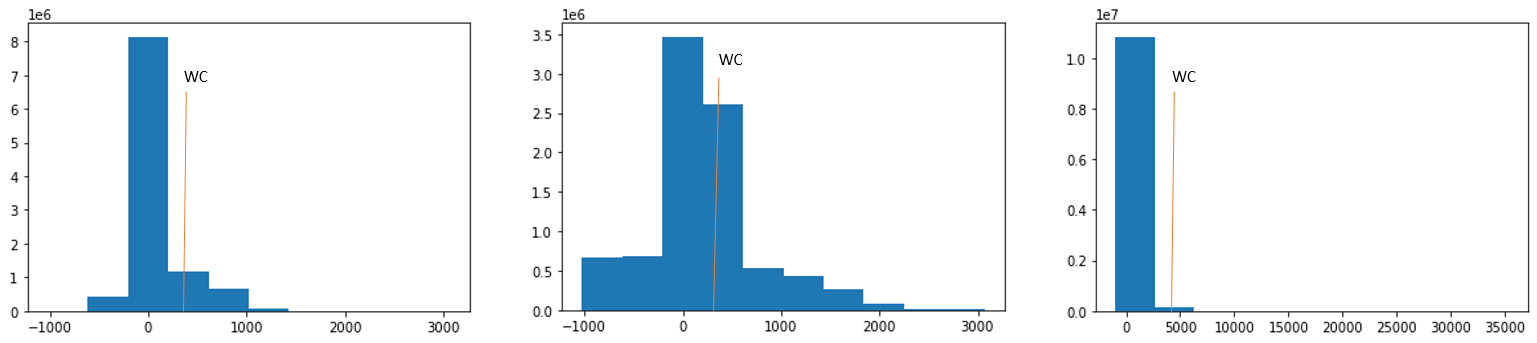}}
\caption{The intensity histograms of the different types of scans. WC represents the Window Centre calculated at run-time based on each histogram.}
\label{fig:f1}
\end{figure*}

\subsection{Data Aquisition}

For this study, 1010 dental CBCT scans were obtained from the PACS of a dental hospital. The data was annotated in two stages; annotation in the first stage was carried out by 28 trained medical students and at the second stage 6 doctors conducted the validation of annotated data. The CBCT scans were in DICOM format with voxel spacing ranging from $0.3mm$ to $0.39mm$. The annotated data was available as a set of floating point polygon coordinates each for the left and right canal separately, stored in JSON file format for each patient. The spatial resolution of scans ranged from 512 $\times$ 512 $\times$ 460 voxels to 670 $\times$ 670 $\times$ 640 voxels. The CBCT scans are divided into three different types based on Hounsfield unit (HU) values ranges. The three different scans have ranges from -1000 to +1000, -1000 to +2000, and 0 to 5000 HU . However, not all dataset is utilized for testing and training. Many experiments were conducted using 100, 200, 300, and 400 scans and tested on 500 samples.

\subsection{Data Pre-processing}
 Although CT scans follow a worldwide standard for ranges of HU values for different body parts like teeth, gums, bones, etc., 3D CBCT scans, on the other hand, follow no such standard and therefore, can have different ranges of HU values and relative intensities when acquired from different manufacturers and under different scanning conditions.
 Fixed window levels and window widths can give varying contrasts and can be a cause for poor results during processing. To make the algorithm more robust dynamic windowing was used for the windowing of scans. This was done by calculating the window levels and widths (WL/WW) on run-time by analyzing the trend of the intensity histogram of the scan to ensure a standard contrast of the scan after windowing.
The intensity histogram of three different types of
CBCT scans, each acquired from a different manufacturer, and the placement of their calculated Window Centre (WC) through the above logic is shown in Figure \ref{fig:f1}.
To perform the dynamic windowing, intensity histogram of the scan was evaluated and the intensity with the highest frequency was set as the window center. The window width (WW) is determined by the range of intensities, the longer the range, the higher the window width with lesser change in WW as the range increases to very high intensities.

\subsection{Network Architecture}
\label{arch}
The main problem in the segmentation of the mandibular canal is the unbalanced numbers of data between the mandibular canal and background classes. This problem often leads to misclassifications, especially at pixels on the boundary of the canal. Another problem that arises while refining the results of segmentation was computational power. Thus, in this study, two cascade networks were utilized to produce a full-resolution segmentation output. The first CNN performed a coarse segmentation of the MC and the second network utilized the VOI from the first network to produce refined segmentation. The output of the first model was used to isolate the left and right parts of the face as well as crop the regions around the mandibular canal in CBCT scan. This gives two VOIs i.e. region around the left mandibular canal and region around the right mandibular canal. These cropped VOIs were used as the input of second model, which produced the fine segmentation of left and right mandibular canals.
 \begin{figure}[h]
\centering
\centerline{\includegraphics[width=.5\textwidth]{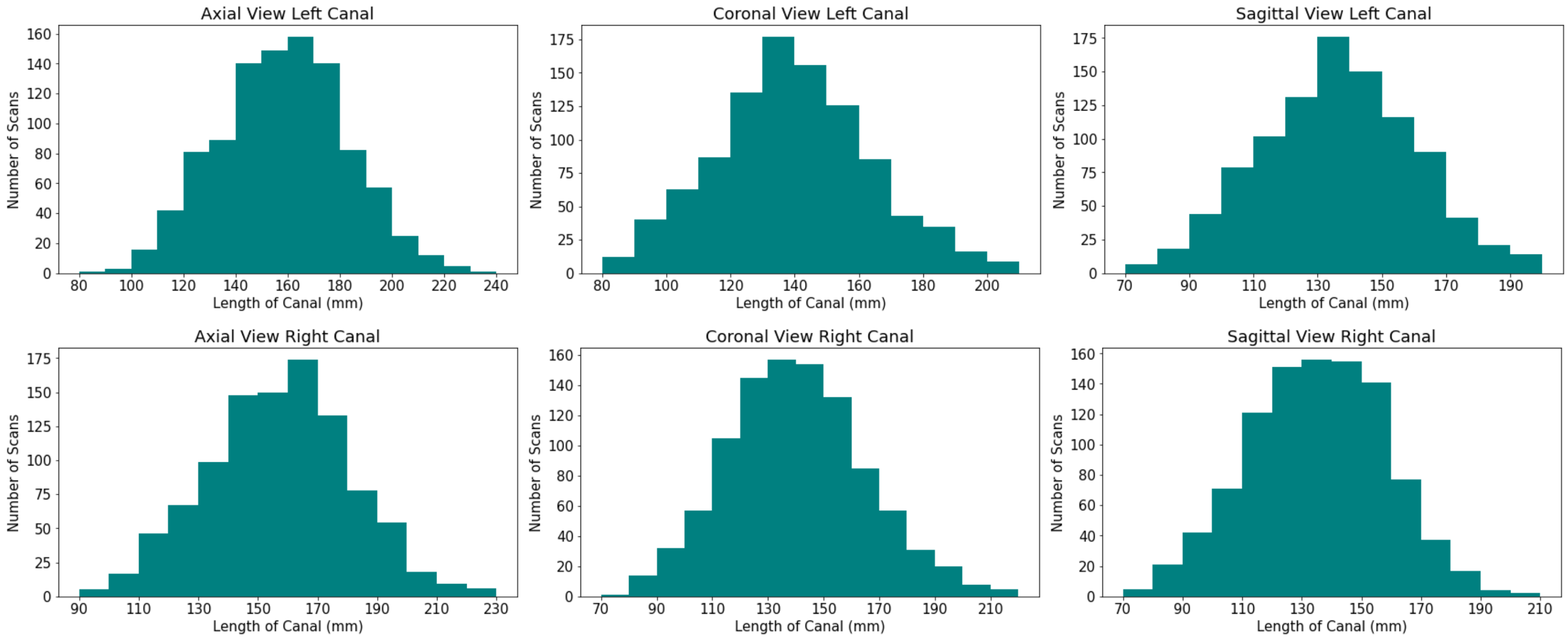}}
\caption{Histogram depicting the difference in size of left and right mandibular canal}
\label{figleft}
\end{figure}
\begin{figure*}[h]
\centering
\centerline{\includegraphics[width=.8\textwidth]{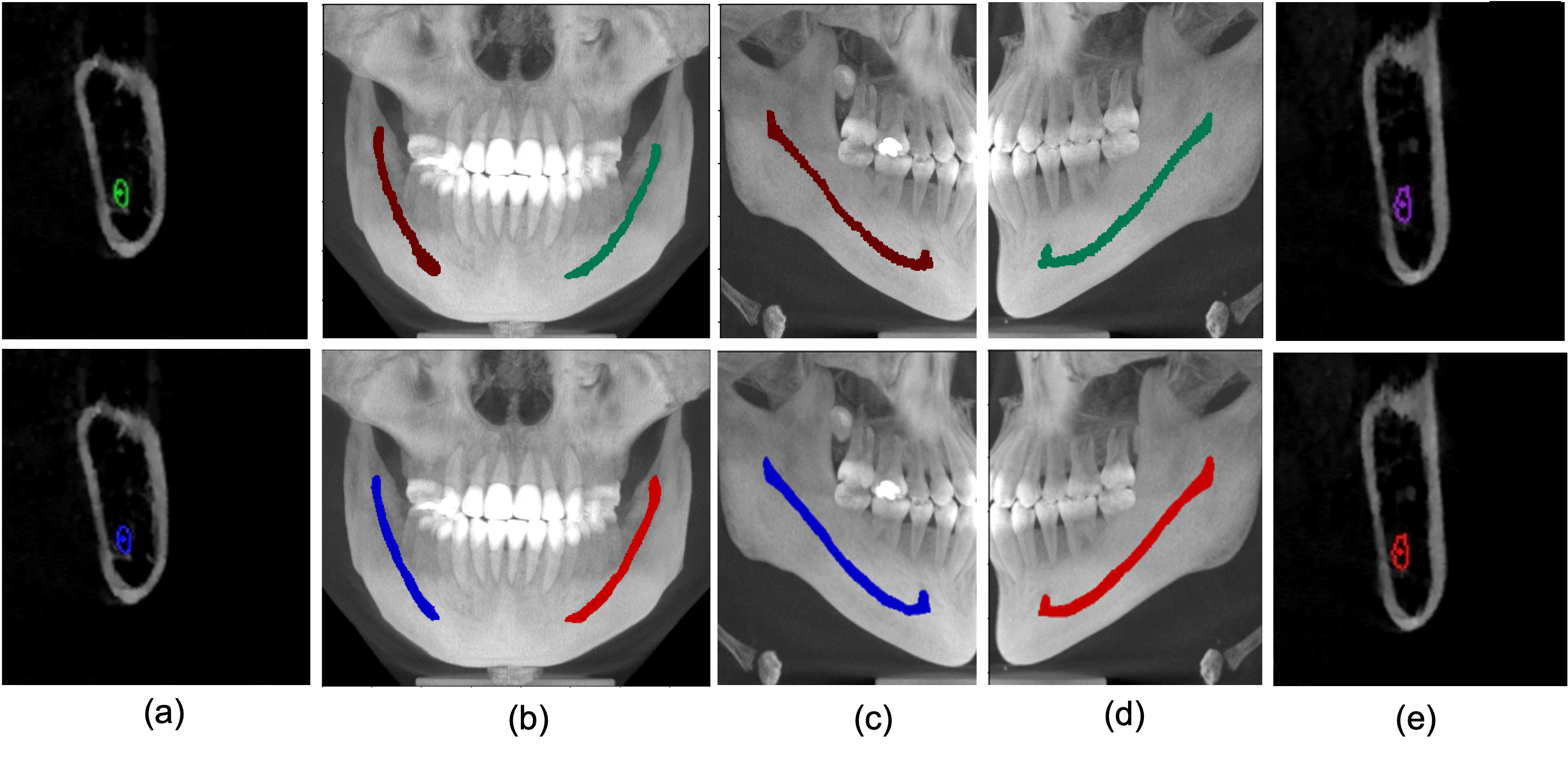}}
\caption{Ground truth (row 1) vs Model Prediction (row 2). a) Parasagittal view of right MC. b) Maximum Intensity Projection of Coronal view (blue as right canal and red as left canal. c) Right canal comparison with ground truth. d) Left canal Comparison with ground truth. e) Parasagittal view of Left Canal }
\label{fig:res}
\end{figure*}
\subsubsection{Jaw Localization}
The main motivation behind this step was the problems faced during segmentation of the mandibular canal in 3D. CBCT scan included the whole face and jaw whereas the region of interest is only the jaw part. This creates a problem of unbalanced numbers of data between object and background classes. This problem often leads to misclassifications, especially on the boundary of canal region. This reduces the accuracy and creates noisy segmentation maps of canal. Moreover, it is very computationally expensive to optimize a model for 1010 scans of size $512\times512\times 1$. Reducing the size of the scan would further affect the appearance of mandibular canal. Thus we utilized 20\% of the data i.e. 200 scans to train a localization model to coarsely segment the mandibular canal. This segmentation is utilized to roughly localize the canal and input to the second model for segmentation.

Since the anatomical contexts in 3D medical pictures are far more complex than those in 2D images, 3D variations of U-Net with significantly more parameters are often needed to capture more representative characteristics. A large number of parameter weights and depth in the 3D U-Net, however, creates a variety of optimization challenges, including over-fitting, slow convergence rate, gradient vanishing, and repetitive computation while training. The vanishing gradient can be described as the decrease in learning of the network in forward propagation due to too many parameters or hidden layers which affects the learning of the network overall. Moreover, there is a lot of redundant information in full-size 3D CBCT scans along with the volume of interest, which distorts the results and optimization takes too much time. 

These issues were resolved by using Deeply Supervised Attention UNet architecture \cite{8983292}. The input to the network
is a 3D CBCT scan $x \in \mathbb{R}^{512 \times 512 \times 1}$, and the output is a segmentation map $\Phi(x) \in[0,1]^{512 \times 512 \times 1}$. The output of the model is a segmentation mask which coarsely segments the canals. 

The network consists of one encoder block and one decoder block. The encoder network learns to extract all the necessary information from the input image, which is then passed to the decoder. Each decoder block consists of attention gate skip connections from the encoder to the decoder. Attention gate assists the model to select more useful features. It takes two inputs; the up-sampling feature in the decoder and the corresponding depth feature in the encoder as shown in the diagram \ref{fig:mod}. The feature from the encoder is used as a gating signal to enhance the learning of feature in the decoder. Attention gates automatically learn to focus on target
structures without additional supervision. At test time, these gates generate soft region proposals implicitly on runtime and highlight salient features useful for a specific task. Moreover, they reduce the computational load and improve the model sensitivity and accuracy for dense label predictions by suppressing feature activations in irrelevant regions.

In order to capture the inter-slice connectivity of the canal and obtain more fine segmentation results, the framework combines the current 3D Attention U-Net model with a 3D deep supervision mechanism during training. In order to strengthen the propagation of gradient flows inside the network and therefore acquire more effective and representative features, a 3D deep supervision method greatly regulates the training of the hidden layers. The deep supervision regularize weights of layers for encoding which have been learned. Deep supervision is only used in training mode as it helps with segmentation by properly regularizing the network weights.

The segmentation masks, produced in this step were used to divide the scan into left and right part and the VOI was extracted for fine segmentation of canal by cropping the region around segmented canal. 

\subsubsection{3D Mandibular Canal Segmentation}

 \label{secpro}

The size of mandibular canal was analyzed statistically as shown in the figures \ref{figleft}. It is visible that both the right canals and left canals vary in size. Thus, to ensure that the performance of model remains similar for all sizes of canal VOIs, Residual UNet architecture with multi-scale inputs was utilized to perform the task of 3D segmentation of mandibular canal. The size of three inputs is kept 208 $ \times $ 240 $ \times $ 240, 144 $ \times $ 176 $ \times $ 176 and 108 $ \times $  144 $ \times $ 144. The benefit of multi-scale input is that it caters to all the different available sizes of the mandibular canal and hence reduces the segmentation error around the boundary. This structure enables the encoder of the network to extract features better. The output was binarized by applying a thresholding of 0.5.

The ResUNet or Deep Residual UNet architecture was utilized for 3D mandibular canal segmentation \cite{article}, an architecture that relies on deep residual learning and U-Net.
Its structure can be divided into an encoding network and a decoding network. The two consecutive layers are applied to the basic residual block and same padding is used in the encoding branch. A batch normalization layer follows each convolutional layer, followed by a ReLU layer (non-linear layer). Downsampling is done by max-pooling operation after the residual block. The number of feature channels is doubled at each down-sampling step. In order to restore the size of the segmented output, the same amount of up-sampling operations are carried out in the decoding network. A transposed convolution is used to achieve each up-sampling, and the number of channels of feature is reduced by half. After passing through the channel attention block, skip connections are created to transfer features from the encoder to the decoder, and basic residual blocks with two successive convolutional layers (with the same padding) are used for feature extraction. Similar to this, a batch normalization layer and a ReLU layer is placed after every convolutional layer. By using concatenation, which is employed in U-Net, the encoded and decoded data are combined.

\begin{figure*}[t]
\centering
\centerline{\includegraphics[width=.7\textwidth]{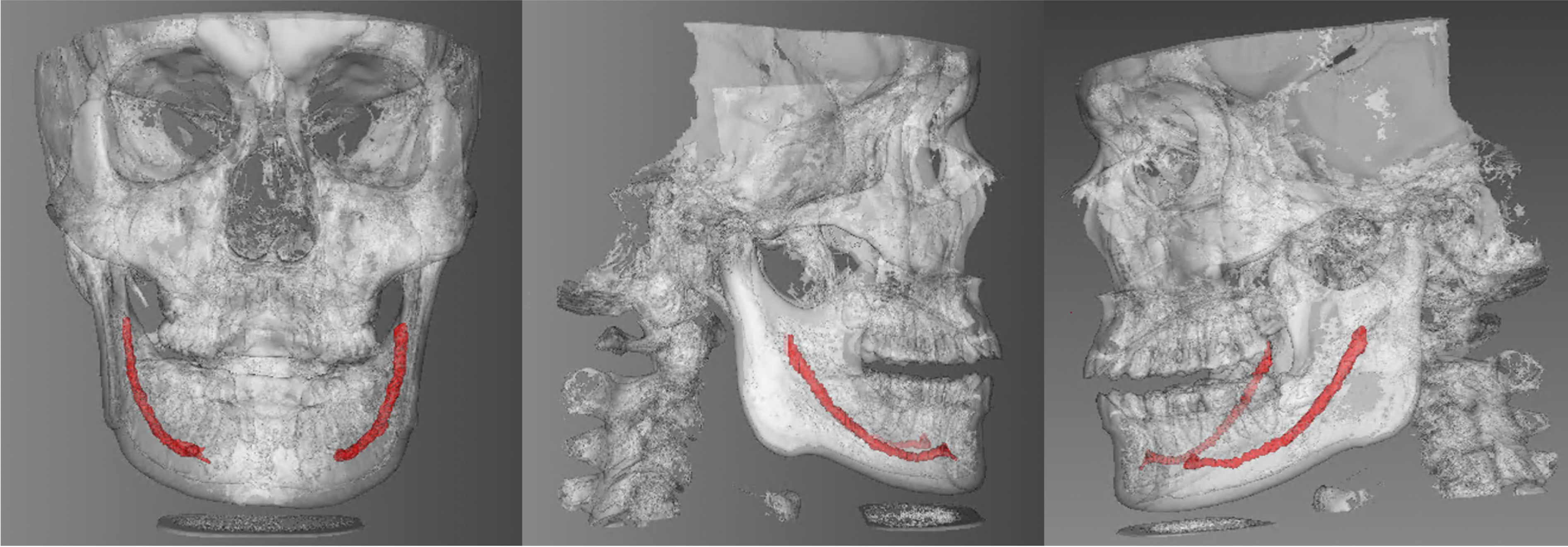}}
\caption{3D rendered mandibular canal predicted by model}
\label{fig:3d}
\end{figure*}

\begin{table*}[h]
\centering
\begin{tabular}{@{}lrrrr@{}}
\toprule
 &
  \multicolumn{1}{l}{\textbf{Without Multiscale}} &
  \multicolumn{1}{l}{\textbf{With Multiscale}} &
  \multicolumn{1}{l}{\textbf{without Resiudual Connections}} &
  \multicolumn{1}{l}{\textbf{with Residual Connections}} \\ \midrule
\textbf{mIOU}       & 0.779 & 0.8   & 0.785 & 0.8   \\
\textbf{Precision}  & 0.683 & 0.69  & 0.679 & 0.69  \\
\textbf{Recall}     & 0.81  & 0.83  & 0.824 & 0.83  \\
\textbf{Dice Score} & 0.72  & 0.748 & 0.72  & 0.748 \\
\textbf{F1 Score}   & 0.741 & 0.749 & 0.745 & 0.749 \\ \bottomrule
\end{tabular}
\caption{Ablation study; comparison of utilized architecture with and without residual connections as well as multiscale inputs.}
\label{tab:ab}
\end{table*}
\begin{table}[h]
\centering
\begin{tabular}{@{}lrrr@{}}
\toprule
\textbf{Evaluation Metric} & \multicolumn{1}{l}{\textbf{Left Canal}} & \multicolumn{1}{l}{\textbf{Right Canal}} & \multicolumn{1}{l}{\textbf{Overall}} \\ \midrule
                           & \multicolumn{1}{l}{\textbf{}}           & \multicolumn{1}{l}{\textbf{}}            & \multicolumn{1}{l}{\textbf{}}        \\
mIOU        & \textbf{0.79} & \textbf{0.8} & \textbf{0.8} \\
Precision   & 0.67          & 0.71         & 0.69         \\
Recall      & 0.84          & 0.81         & 0.83         \\
Dice Score  & 0.74          & 0.75         & 0.748        \\
F1 Score    & 0.74          & 0.75         & 0.749        \\
Specificity & 0.99          & 0.99         & 0.99         \\ \bottomrule
\end{tabular}
\caption{Evaluation of proposed architecture on different metrices.}
\label{tab:q}
\end{table}
\subsubsection{Post-processing}
The outputs of the network were manually analyzed and it was concluded that a few canals were not fully connected, that is, there were splits between the canals. The noise was also observed in some scans near the canal. Thus, a post-processing method is developed to make improvements in the segmented canal. To develop such methods, simple image processing techniques such as dilation, erosion, opening, and closing morphological operations were performed. This method removes the noise as well as ensures that the canal is connected.

\begin{table*}[h]
\begin{tabular}{@{}lrrrrrrrr@{}}
\toprule
\textbf{Number of scans} & \multicolumn{2}{r}{\textbf{100}} & \multicolumn{2}{r}{\textbf{200}} & \multicolumn{2}{r}{\textbf{300}} & \multicolumn{2}{r}{\textbf{400}} \\
 & \multicolumn{1}{l}{\textbf{Left Canal}} & \multicolumn{1}{l}{\textbf{Right Canal}} & \multicolumn{1}{l}{\textbf{Left Canal}} & \multicolumn{1}{l}{\textbf{Right Canal}} & \multicolumn{1}{l}{\textbf{Left Canal}} & \multicolumn{1}{l}{\textbf{Right Canal}} & \multicolumn{1}{l}{\textbf{Left Canal}} & \multicolumn{1}{l}{\textbf{Right Canal}} \\ \midrule
mIOU & \textbf{0.765} & \textbf{0.771} & \textbf{0.78} & 0.789 & 0.79 & 0.8 & 0.798 & 0.806 \\
Precision & 0.639 & 0.667 & 0.65 & 0.69 & 0.67 & 0.71 & 0.69 & 0.72 \\
Recall & 0.818 & 0.795 & 0.832 & 0.8 & 0.84 & 0.81 & 0.854 & 0.819 \\
Dice Score & 0.721 & 0.731 & 0.734 & 0.746 & 0.741 & 0.753 & 0.746 & 0.759 \\
F1 Score & 0.72 & 0.73 & 0.73 & 0.74 & 0.75 & 0.76 & 0.76 & 0.77 \\ \bottomrule
\end{tabular}
\caption{Comparison of model performance with different number of training samples (100, 200, 300 and 400).}
\label{tab:com}
\end{table*}

\begin{table*}[h]
\centering
\begin{tabular}{@{}llrrrrrl@{}}
\toprule
\textbf{Study} &
  \textbf{Source of Data} &
  \multicolumn{1}{l}{\textbf{Training Scans}} &
  \multicolumn{1}{l}{\textbf{mIOU}} &
  \multicolumn{1}{l}{\textbf{Precision}} &
  \multicolumn{1}{l}{\textbf{Recall}} &
  \multicolumn{1}{l}{\textbf{Dice Score}} &
  \textbf{F1 Score} \\ \midrule
 &
   &
  \multicolumn{1}{l}{} &
  \multicolumn{1}{l}{} &
  \multicolumn{1}{l}{} &
  \multicolumn{1}{l}{} &
  \multicolumn{1}{l}{} &
   \\
\cellcolor[HTML]{FFFFFF}Jaskari et al., 2020 &
  Private &
  457 &
  \multicolumn{1}{l}{-} &
  \multicolumn{1}{l}{-} &
  \multicolumn{1}{l}{-} &
  0.575 &
  \multicolumn{1}{r}{0.575} \\
\cellcolor[HTML]{FFFFFF}Kwak et al., 2020 (3D) &
  Private &
  61 &
  0.959 &
  \multicolumn{1}{l}{-} &
  \multicolumn{1}{l}{-} &
  \multicolumn{1}{l}{-} &
  - \\
Dhar et al., 2021 &
  Private &
  157 &
  0.7 &
  0.63 &
  0.51 &
  \multicolumn{1}{l}{-} &
  \multicolumn{1}{r}{0.575} \\
\cellcolor[HTML]{FFFFFF}{\color[HTML]{222222} Verhelst et al., 2021} &
  Private &
  190 &
  0.945 &
  0.952 &
  0.993 &
  0.972 &
  - \\
Lahoud et al., 2022 &
  Private &
  166 &
  0.636 &
  0.782 &
  0.792 &
  0.774 &
  - \\
our method &
  Private &
  100 &
  0.768 &
  0.653 &
  0.807 &
  0.729 &
  \multicolumn{1}{r}{0.725} \\
our method &
  Private &
  200 &
  0.785 &
  0.67 &
  0.816 &
  0.74 &
  \multicolumn{1}{r}{0.735} \\
our method &
  Private &
  300 &
  0.795 &
  0.69 &
  0.832 &
  0.75 &
  \multicolumn{1}{r}{0.755} \\ \bottomrule

\end{tabular}
\caption{Comparison of our proposed architecture with other studies.}
\label{tab:os}
\end{table*}
\subsection{Training of AI model}
For this study, all the CNN architectures were implemented using the Keras framework \cite{chollet2015keras} with TensorFlow \cite{tensorflow2015-whitepaper} as back-end. We performed
our experiment on two powerful
NVIDIA Titan RTX GPUs with 4608 CUDA cores and 24GB GDDR6 SDRAM. The batch size for this experiment was set to 2 for both models, and the proposed architecture
is optimized by Adam optimizer. \num{1e-5} is the learning rate set to train the model. To reduce the training
time, and use the GPU efficiently, we use 10 to 20 percent of the 3D CBCT scans for training of model for jaw localization as well as canal segmentation. The size of images while training the jaw localization model is kept to 512 x 512 x512. After localizing the jaw, the 3D images are cropped and resized to three fixed sizes as mentioned in the section \ref{secpro}. We used the dice loss function equation \ref{eqdice} to calculate the loss. Labels are the segmentation annotation of images containing 0 as background and 1 as foreground. We trained the localization model for 50 epochs and the 3D segmentation models for 80 epochs by keeping the learning rate lower in order to train a generalized model.
The batch size, epoch, and learning rate were reset depending
upon the need. 
\begin{equation}
\begin{aligned}
\text { Dice Loss} &=1- \text {Dice Coefficient}
\end{aligned}
\label{eqdice}
\end{equation}
where, dice coefficient is given by equation \ref{dice}.
\subsection{Performance Measures}
In order to measure the performance of the deep learning model, we calculated the dice score, mean IOU, precision, recall, F1 score and specificity using the following equations:
\begin{equation}
\begin{aligned}
\text { precision } &=\frac{T P}{T P+F P}
\end{aligned}
\end{equation}
\begin{equation}
\begin{aligned}
\text { recall } &=\frac{T P}{T P+F N}
\end{aligned}
\end{equation}

\begin{equation}
\begin{aligned}
F 1 score &=\frac{T P}{T P+0.5(F P+F N)}
\end{aligned}
\end{equation}

\begin{equation}
\begin{aligned}
I O U &=\frac{T P}{T P+F P+F N}
\end{aligned}
\end{equation}

\begin{equation}
\mathrm{DSC}=\frac{2 \mathrm{TP}}{2 \mathrm{TP}+\mathrm{FP}+\mathrm{FN}}
\label{dice}
\end{equation}

Where, TP refers to true Positives, FP refers to False Positives, FN refers to False Negatives, and TN refers to True Negatives. IOU refers to Intersection Over Union and DSC is Dice Coefficient.
\section{Results}
\label{res}
\subsection{Ablation Study}
In this section, we conduct comprehensive experiments
to analyze the effectiveness of our utilized 3D segmentation model on 300 scans.
Table \ref{tab:ab} shows the model evaluation with and without multi-scale model, as well as with and without residual connections. It can be concluded from our experiments that including multi-scale inputs can improve dice score up to 2.8 $\%$, moreover, the residual connections also help increase the performance by 1.1 $\%$. Multi-scale inputs helps the model to learn varying size of mandibular canals thus improving the performance. It also assists the model to produce more smooth boundary. Moreover, an improvement in the boundary of mandibular canal was observed when model was trained with residual connections as compared to without residual connections.

\subsection{Qualitative Results}
Figure \ref{fig:res} shows the qualitative results with the ground truth in figures (a) and output of the model in figures (b). The rendering of canal, predicted by the model is also shown in figure \ref{fig:3d}.

\subsection{Effect of amount of data on results}
Originally, there were total 1000 scans available for the experiment. However, training two models on this huge amount requires a lot of time and computation power. For each 100 scans, the model took 2.68 hours to complete each epoch. Thus to avoid that, we analyzed the effect of increasing the amount of training data on model performance. We trained the model on 100, 200, 300 and 400 scans. It was concluded that initially, model's performance improved by 1.3 $\%$ but as the amount of data was increased further, the change is performance was negligible i.e 0.7 $\%$, 0.5 $\%$ respectively. We selected the model trained on 300 scans for our final results. Table \ref{tab:com} shows the effect of of increasing the data on performance of model.

\subsection{Quantitative Results}

 The dice score, precision, and recall give us insight into the voxel-level segmentation performance of the model. Table \ref{tab:q} shows the performance of the utilized technique on left canal, right canal as well as combined results. The number of scans for the reported results are 300. The metrics we chose to evaluate our model includes dice score, mean intersection over union, F1 score and specificity. Our model performed very well with the overall mean IOU score of 80$\%$ and dice score of 75$\%$. It is to be noted that the segmentation dice of left canal is slightly less than the right canal and this trend can be seen in all the conducted experimentation.

\subsection{Comparison with the previous techniques}
The dataset used for the research is different from other research papers hence no direct comparison can be drawn. However, we draw a comparison based on the model utilized. Table 
shows the comparison between our technique and previously proposed techniques. 

\section{Analysis and Discussion}
\label{disc}
To prevent any surgical complications, a correct identification of the mandibular canal anatomy is an essential element of the both preoperative and postoperative planning of third molar extractions and implant surgeries. But for a variety of reasons, it is anticipated to be a labor- and time-intensive process to accurately identify the complete canal structure. The most used 3D dental imaging method, CBCT, has less contrast than CT, which makes it difficult to manually distinguish mandibular canals. The structural continuity of MC segmentation in CBCT images is thus impacted by the limited visibility of MCs, such as in ambiguous or unclear cortical bone regions. Variance of the cortications and bone densities of the canal wall, variation in travel routes of the canal, and the distribution of vessels and nerve branches all contribute to the low visibility of the MC itself.
These defects, make the task of annotating the mandibular canal very challenging for radiologists and the results become prone to human error. Application of deep learning algorithms can reduce the complexity of task and assist the radiologist in reducing the burden.  Moreover, it can reduce the time consumed and amount of labour utilized by making the annotation process simple and easy. However, these networks shows some errors in detection of canals especially boundary region. Thus we overcame the problem by using residual connections as well as multi-scale inputs in the architecture. By conducting several experimentation and validation of results, it was concluded that the model is slightly less accurate in the segmentation of left mandibular canal as compared to right mandibular canal. The network also outputs some noise along with canal output and the canal was disconnected in some scans.

\section{Conclusion}
\label{co}
In this paper, we propose a novel dual-stage deep learning based scheme for automatic detection of mandibular canal. We first enhance the CBCT scans by employing the novel histogram-based dynamic windowing scheme which improves the visibility of mandibular canals. After enhancement, we design 3D deeply supervised attention U-Net architecture for localizing the mandibular canals within the volumes of interest (VOIs). Finally, each VOIs is fed to multi-scale input residual U-Net architecture to accurately segment the mandibular canals. The proposed method has been rigorously evaluated on 500 scans and an extensive quantitative as well as qualitative/visual analysis has been performed. The results demonstrate that our framework achieves improved segmentation performance compared to previous state-of-the-art methods. Particularly, our scheme obtained 0.748 and 0.8 values of the average dice score and mean intersection over union on test dataset, respectively. Future work includes reducing the number of stages utilized as this method requires training of multiple models which leads to high computation cost and time.

\end{document}